\documentstyle[aps,epsf,floats]{revtex}

\begin{document}
\parskip 2mm
\twocolumn[\hsize\textwidth\columnwidth\hsize\csname@twocolumnfalse%
\endcsname
\title{Pair contact process with diffusion -- \\
       A new type of nonequilibrium critical behavior?}
\author{Haye Hinrichsen\\[2mm]}

\address{Theoretische Physik, Fachbereich 10,\\
     Gerhard-Mercator-Universit{\"a}t Duisburg,
     47048 Duisburg, Germany}

\date{September 20, 2000, to appear in Phys. Rev. E}
\maketitle
\begin{abstract}
In the preceding article Carlon {\it et al.} investigate the 
critical behavior of the pair contact process with 
diffusion. Using density matrix renormalization group methods, they 
estimate the critical exponents, raising the possibility that 
the transition might belong to the same universality 
class as branching annihilating random 
walks with even numbers of offspring. This is surprising
since the model does not have an explicit parity-conserving symmetry. 
In order to understand this contradiction, we
estimate the critical exponents by Monte Carlo simulations. 
The results suggest that the transition might belong to a different 
universality class that has not been investigated before.
\end{abstract}

\pacs{{\bf PACS numbers:} 05.70.Ln, 64.60.Ak, 64.60.Ht}]
%


Symmetries and conservation laws are known to
play an important role in the theory of
nonequilibrium critical phenomena~\cite{MarroDickman99}. 
As in equilibrium statistical mechanics, most phase transitions 
far from equilibrium are characterized by certain universal 
properties. The number of possible universality classes, 
especially in 1+1 dimensions, is believed to be finite. 
Typically each of these universality classes is associated 
with certain symmetry properties. 

One of the most prominent universality classes of nonequilibrium phase 
transitions is directed percolation (DP)~\cite{Kinzel83,Hinrichsen00}. 
According to a conjecture by Janssen and
Grassberger, any phase transition from a fluctuating phase
into a {\em single} absorbing state in a homogeneous system with
short-range interactions should belong to the DP universality
class, provided that there are no special attributes such 
as quenched disorder, additional conservation laws, or unconventional
symmetries~\cite{Janssen81,Grassberger82}. Roughly speaking, the
DP class covers all models following the reaction-diffusion
scheme $A \leftrightarrow 2A$, $A \rightarrow \emptyset$. Regarding
systems with a single absorbing state the DP conjecture is
well established nowadays. However, even various systems 
with infinitely many absorbing states have been found to belong 
to the DP class as well~\cite{JensenDickman93b,MDHM94,MGDL96}.

Exceptions from DP are usually observed if one of the conditions 
listed in the DP conjecture is violated. This happens, for instance, in
models with additional symmetries. An important example is the
so-called parity-conserving (PC) universality class, which is
represented most prominently by branching annihilating random 
walks with two offspring 
$A \rightarrow 3A, \, 2A \rightarrow 
\emptyset$~\cite{TakayasuTretyakov92,ALR93,CardyTauber96}. 
In 1+1 dimensions this process can be interpreted as
as a $Z_2$-symmetric spreading process with branching-annihilating 
kinks between oppositely oriented absorbing domains.
Examples include certain kinetic Ising 
models~\cite{Menyhard94}, interacting monomer-dimer
models~\cite{KimPark94}, as well as generalized versions of the 
Domany-Kinzel model and the contact process with two symmetric absorbing 
states~\cite{Hinrichsen97}.

A very interesting model, which is studied in the present
work, is the (1+1)-dimensional pair contact process (PCP)
$2A\rightarrow 3A$, $2A\rightarrow \emptyset$~\cite{Jensen93a}. Depending
on the rate for offspring production, this model displays a
nonequilibrium transition from an active into an inactive phase.
Without diffusion the PCP has infinitely many absorbing states and
the transition is found to belong to the universality class of DP. 
The pair contact process with diffusion (PCPD), however, 
is characterized by a different type of critical behavior. 
In the inactive phase, for example, the order parameter 
no longer decays exponentially, instead it is governed by 
an annihilating random walk with an algebraic decay. 
Moreover, the PCPD has only two absorbing states, namely, the empty 
lattice and the state with a single diffusing particle.
For these reasons the transition is expected to cross 
over to a different universality class. The PCPD, also called
the annihilation/fission process, was first 
proposed  by Howard and T{\"a}uber~\cite{HowardTauber97}
as a model interpolating between ``real'' and ``imaginary'' noise. 
Based on a field-theoretic renormalization group study,
they predicted non-DP critical behavior at the transition.

In the preceding article, Carlon, Henkel, and 
Schollw\"ock~\cite{CHS99} investigate a lattice model of the 
PCPD with random-sequential 
updates. In contrast to Ref.~\cite{HowardTauber97}, each 
site of the lattice can be occupied by at most one particle, 
leading to a well-defined particle density 
in the active phase. Performing a careful density matrix 
renormalization group (DMRG) study~\cite{PWKH99,CHS99DMRG}, 
Carlon {\it et~al.} estimate two of four independent
critical exponents.
Depending on the diffusion rate $d$, their estimates for 
$\theta=z$ vary in the range $1.60(5) \ldots 1.87(3)$
while $\beta/\nu_\perp$ is found to be close to $0.5$.
Since these values are close the the PC exponents 
$z=1.749(5)$ and $\beta/\nu_\perp=0.499(2)$, they suggest
that the transition might belong to the PC universality class.
\begin{figure}
\epsfxsize=85mm
\centerline{\epsffile{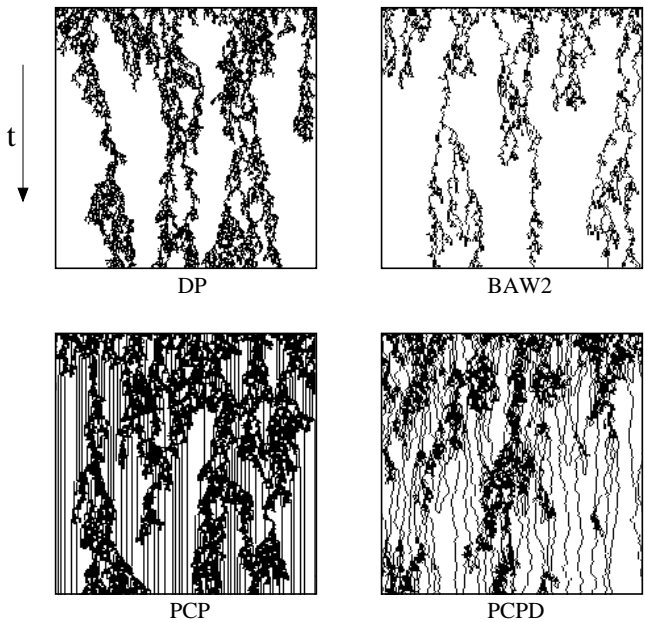}}
\vspace{2mm}
\caption{
\label{FIGDEMO}
Typical space-time trajectories of four different critical 
nonequilibrium processes
starting with a fully occupied lattice.
The four panels show directed percolation (DP), 
a branching annihilating random walk with two offspring (BAW2), 
and the pair contact process without (PCP) and with diffusion (PCPD).
}
\end{figure}

The conjectured PC transition poses a puzzle. 
In all cases investigated so far, the PC
class requires an {\it exact} symmetry on the level
of microscopic rules. In 1+1 dimensions this symmetry may be
realized either as a parity conservation law or as an explicit
$Z_2$ symmetry relating two absorbing states. 
In the PCPD, however, the dynamic rules are neither
parity conserving nor invariant under an obvious symmetry
transformation. Yet how can the critical properties of the transition
change without introducing or breaking a symmetry?
As a possible way out, there could be a hidden
symmetry in the model, but we have good reasons 
to believe that there is no such hidden symmetry
or conservation law in the PCPD. This would imply 
that the PC class is not characterized by a 
``hard'' $Z_2$ symmetry on the microscopic level, rather, 
it may be sufficient to have a ``soft''  equivalence of 
two different absorbing states in the sense that 
they are reached by the dynamics with the same probability. 

In this paper I suggest that the transition in the PCPD 
might belong to a different yet unknown universality class.
The reasoning is based on the conservative point of view that
a ``soft'' equivalence between two absorbing states is not 
sufficient to obtain PC critical behavior. As described in
Ref.~\cite{Hinrichsen97}, the essence of the PC class is a 
competition between two types of absorbing domain that are 
related by an {\it exact} $Z_2$ symmetry. Close to criticality
these growing domains are separated by {\em localized} regions of
activity. In 1+1 dimensions, these active regions may be interpreted
as kinks between oppositely oriented domains, which, by their very
nature, perform an unbiased parity-conserving branching-annihilating
random walk. In the PCPD, however, it is impossible to give an exact 
definition of ``absorbing domains.'' 
We can, of course, consider empty intervals
without particles as absorbing domains. 
Yet, what is the meaning of a domain
with only one diffusing particle? And even if such a definition were
meaningful, what would be the boundary between an 
empty and a ``single-particle'' domain? Moreover, in PC models there
are two separate sectors of the dynamics (namely, with even and
odd particle numbers), whereas there are no such sectors in the PCPD.
In fact, even when looking at typical space-time trajectories, 
the PCPD differs significantly from a standard branching-annihilating 
random walk with two offspring 
(see Fig.~\ref{FIGDEMO}). In particular, offspring production in the 
PCPD occurs spontaneously in the bulk when two
diffusing particles meet, whereas a branching-annihilating
random walk generates offspring all along 
the particle trajectories. Therefore, it is 
reasonable to expect that the two critical phenomena are not fully
equivalent.

\begin{figure}
\epsfxsize=85mm
\centerline{\epsffile{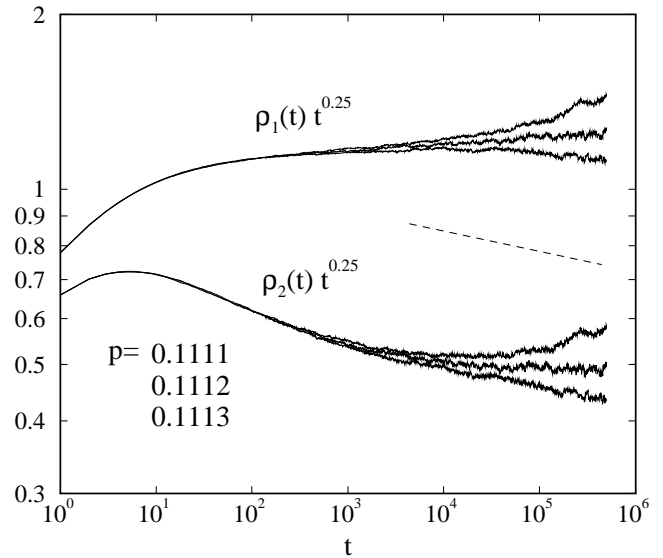}}
\caption{
\label{FIGDYN}
The density of particles $\rho_{1}(t)$  and the
density of pairs $\rho_{2}(t)$ times $t^{0.25}$
as a function of time measured in Monte Carlo steps. 
Upward (downward) curvature indicates
deviations from the critical point. The dashed line represents
the slope corresponding to the PC exponent $\delta \simeq 0.285$.
}
\end{figure}

In order to investigate this question in more detail,
it is useful to compare the DMRG estimates with
numerical results obtained by Monte Carlo simulations. 
It is important to note that there 
are two possible order parameters, namely, the particle density
\begin{equation}
\rho_{1}(t) = \frac{1}{L} \sum_i s_i(t) 
\end{equation}
and the density of {\em pairs} of particles
\begin{equation}
\rho_{2}(t) = \frac{1}{L} \sum_i s_i(t) s_{i+1}(t) ,
\end{equation}
where $L$ is the system size and
$s_i(t)=0,1$ denotes the state of site $i$ at time $t$. 
Performing high-precision simulations it turns out 
that the critical behavior at the transition
is characterized by unusually strong corrections to 
scaling~\cite{Simulations}.
These scaling corrections are demonstrated
in Fig.~\ref{FIGDYN}, where the temporal decay of the two
order parameters for $d=0.1$ is shown as a function of time 
running up to almost $10^6$ time steps. The pronounced curvature of 
the data in the double-logarithmic plot demonstrates 
the presence of strong corrections to scaling. 
Interestingly, the two curves bend in opposite directions
and tend toward the same slope.
Thus, in contrast to the mean-field prediction, 
$\rho_1(t)$ and $\rho_2(t)$ seem
to scale with the same exponent. 
Discriminating between the negative 
curvature of $\rho_1(t)$ and the positive curvature of $\rho_2(t)$,
we estimate the critical point and the exponent 
$\delta=\beta/\nu_\parallel$ by
\begin{equation}
\label{Estimates}
p_c=0.1112(1) , \qquad \delta=\beta/\nu_\parallel=0.25(2).
\end{equation}

While this estimate deviates only slightly from the known PC 
value $0.286(2)$, other exponents deviate more significantly. 
Performing dynamic simulations starting with a single pair 
of particles~\cite{GrassbergerTorre79}, we measure 
the survival probability $P(t)$  that the system has not yet 
reached one of the two absorbing states~\cite{Remark}, 
the average number of particles $N_1(t)$ and pairs $N_2(t)$, 
and the mean square spreading from the origin $R^2(t)$ 
averaged over the surviving runs. At criticality, these 
quantities should obey asymptotic power laws, 
$P(t) \sim t^{-\delta'}$, $N_1(t)\sim N_2(t) \sim t^{\eta}$, and
$R^2(t) \sim t^{2/z}$,
with certain dynamical exponents $\delta'$ und $\eta$. Notice that
in non-DP spreading processes the two exponents 
$\delta=\beta/\nu_\parallel$ and $\delta'=\beta'/\nu_\parallel$ 
may be different. Going up to $2\times 10^5$ time steps we obtain
the estimates
\begin{equation}
\delta'=0.13(2), \quad \eta=0.13(3), \quad z=1.83(5)\,.
\end{equation}
Although the precision of these simulations is only moderate, the
estimates differ significantly from the PC exponents
$\delta'=0.286$, $\eta=0$ in the even sector and 
$\delta'=0$, $\eta=0.285$ in the odd sector.
The exponent $z$, on the other hand, seems to be close to the
PC value $1.75$.

The most striking deviation is observed in the exponent $\beta$,
which is not accessible in DMRG studies. Here the estimates
seem to decrease with increasing numerical effort. As an upper
bound we find
\begin{equation}
\beta < 0.67.
\end{equation}
Even more recently, \'Odor studied a slightly 
different version of the PCPD on a parallel computer, reporting 
the estimate $\beta=0.58(1)$~\cite{Odor00} which is incompatible 
with the PC exponent $\beta=0.92(2)$.

In summary the critical behavior of the PCPD is 
affected by strong corrections to scaling,
wherefore it is extremely difficult to estimate the critical
exponents. Although DMRG estimates presented in~\cite{CHS99} are
very accurate, they have to be taken with care
since they are affected by scaling corrections as well.
Thus, the apparent coincidence with the exponents of the PC class 
may be accidental. Comparing other exponents, in particular 
the density exponent $\beta$ and the cluster exponents 
$\delta'$ and $\eta$, the PC hypothesis can be ruled out. 

\noindent
I would like to thank J. Cardy, E. Carlon, P. Grassberger, 
M. Henkel, M. Howard, J. F. F. Mendes,
G. {\'O}dor, U. Schollw{\"o}ck, U. T\"auber, and
F. van Wijland for stimulating discussions.


\end{document}